
\input harvmac
\input tables
\def\nt{{\nu_\tau}}
\def\nm{{\nu_\mu}}
\def\ne{{\nu_e}}
\def\ra{\rightarrow}
\def\bell{\bar\ell}
\def\gsim{\ \rlap{\raise 2pt \hbox{$>$}}{\lower 2pt \hbox{$\sim$}}\ }
\def\lsim{\ \rlap{\raise 2pt \hbox{$<$}}{\lower 2pt \hbox{$\sim$}}\ }
\def\l{\lambda}
\def\lp{\lambda^\prime}
\def\ld{\lambda^{\prime\prime}}
\def\H{{\cal H}}
\def\O{{\cal O}}
\def\mod{{\rm mod}}
\def\tm{{\tilde m}}
\def\Ho{{H_{1\alpha}}}
\def\Ht{{H_{2\alpha}}}

\Title{hep-ph/9505248, RU-24-95, WIS-95/21/May-PH}
{\vbox{\centerline{Supersymmetry without R-Parity}
 \centerline{and without Lepton Number}}}
\bigskip
\centerline{Tom Banks}
\smallskip
\centerline{\it Department of Physics and Astronomy}
\centerline{\it Rutgers University, Piscataway, NJ 08855-0849}
\bigskip
\centerline{Yuval Grossman, Enrico Nardi and Yosef Nir}
\smallskip
\centerline{\it Department of Particle Physics}
\centerline{\it Weizmann Institute of Science, Rehovot 76100, Israel}
\bigskip
\baselineskip 18pt

\noindent
We investigate Supersymmetric models where neither R parity nor
lepton number is imposed. Neutrino masses can be kept highly
suppressed compared to the electroweak scale if the $\mu$-terms
in the superpotential are aligned with the SUSY-breaking bilinear
$B$-terms. This situation arises naturally in the framework of
horizontal symmetries. The same symmetries suppress the trilinear
R parity violating terms in the superpotential to an acceptable level.

\Date{5/95}

\newsec{Introduction}
Baryon and lepton number conservation are relics of the ancient history
of particle physics. We know today that they are not likely to be exactly
preserved symmetries of nature. Nonetheless much of the modern
discussion of Supersymmetric models is cast within a framework in which
symmetries that guarantee baryon and lepton number conservation at the
level of renormalizable interactions are assumed
\ref\DRW{S. Dimopoulos, S. Raby and F. Wilczek,
 Phys. Lett. B112 (1982) 133.}
\ref\FIQ{A. Font, L. Ib\'a\~nez and F. Quevedo,
 Phys. Lett. B228 (1989) 79.}
\ref\IbRo{L. Ib\'a\~nez and G.G. Ross, Nucl. Phys. B368 (1992) 3.}
\ref\HiKa{I. Hinchliffe and T. Kaeding, Phys. Rev. D47 (1993) 279.}
\ref\Whit{P.L. White, Nucl. Phys. B403 (1993) 141.}.
The purpose of the present paper is to show that a much larger spectrum
of models may be consistent with the data. Furthermore, since we suspect
that the intricate structure of the quark mass matrix is probably
connected to a horizontal symmetry group, we find it natural to suppose
that this same symmetry group may have something to do with the absence
of what are usually called baryon and lepton number violating processes.

Within the Standard Model, lepton number violating observables
and lepton flavor
changing processes are forbidden because $U(1)_e\times U(1)_\mu\times
U(1)_\tau$ is an accidental symmetry of the (renormalizable)
Standard Model Lagrangian. This makes such processes particularly
sensitive probes of new physics at high energy scales. Thus,
measurements of lepton number violating observables such as
neutrino (Majorana) masses
\ref\PDG{The Particle Data Group, Phys. Rev. D50 (1994) 1173.}
\ref\Aleph{ALEPH: talk presented at Aspen Winter Conference (1995).},
\eqn\numass{
m_\ne\leq\ 5.1\ eV,\ \
m_\nm\leq\ 160\ keV,\ \
m_\nt\leq\ 24\ MeV,}
and lepton flavor changing decays such as \PDG\
\eqn\LFV{
BR(\mu\ra e\gamma)\leq\ 4.9\times10^{-11},\ \
BR(\mu\ra eee)\leq\ 1.0\times10^{-12},}
put severe constraints on extensions of the Standard Model.

Generic Supersymmetric extensions of the Standard Model predict
large contributions to neutrino masses and to lepton flavor
violating decays:
\item{(i)} Sneutrino VEVs give neutrino masses by mixing
neutrinos with the zino $\tilde z$
\ref\nuvevsa{C.S. Aulakh and R.N. Mohapatra,
 Phys. Lett. 119B (1982) 136.}
\ref\nuvevsb{G. Ross and J. Valle, Phys. Lett. 151B (1985) 375.}
\ref\nuvevsc{J. Ellis {\it et. al.} , Phys. Lett.  150B (1985) 142.}
\ref\nuvevsd{D.E. Brahm, L.J. Hall and S. Hsu,
 Phys. Rev. D42  (1990) 1860.}.
\item{(ii)} Quadratic terms (``$\mu$-terms") in the superpotential
give neutrino masses by mixing neutrinos with the (up-)Higgsino
$\tilde\phi_u^0$
\ref\HaSu{ L.J. Hall and M. Suzuki, Nucl. Phys. B231 (1984) 419.}
 \ref\Da{S. Dawson, Nucl. Phys. B261 (1985) 297.}.
\item{(iii)} Trilinear terms in the superpotential induce
tree-level slepton-mediated decays such as $\mu\ra3e$
\HiKa \HaSu \Da
\ref\BGH{V. Barger, G.F. Giudice and T. Han,
 Phys. Rev. D40 (1989) 2987.}.

The bounds \numass\ and \LFV\ severely constrain the Supersymmetric
parameters. Taking $m_{\tilde z}\sim m_Z$, we find
\eqn\boundVEV{\vev{\tilde\nu_\tau}\lsim\sqrt{m_\nt m_{\tilde z}}
\lsim1\ GeV.}
Taking  $m_{\tilde\phi_u^0}\sim m_Z$, we find
\eqn\boundmu{\mu_{\nt\phi_u}\lsim\sqrt{m_\nt m_{\tilde\phi_u^0}}
\lsim1\ GeV.}
Both \boundVEV\ and \boundmu\ become stronger by three orders
of magnitude, namely $\vev{\tilde\nt}\lsim1\ MeV$ and
$\mu_{\nt\phi_u}\lsim1\ MeV$, if the cosmological bound
on long-lived neutrinos, $m(\nu_i)\leq\O(10\ eV)$, holds.
Taking the slepton mass $m_{\tilde\ell}\sim m_Z$,
the bound \LFV\ on $\mu\ra3e$ constrains the product of two
lepton number violating couplings to be
\eqn\boundlam{\l_{k12}\l_{k11}\lsim10^{-6}.}
(We do not consider here the stronger  constraints from baryogenesis
\ref\FuYa{M. Fukugita and T. Yanagida, Phys. Rev. D42 (1990) 1285.}
\ref\FGLP{W. Fishler {\it et. al.} , Phys. Lett. 258B (1991) 45.}
\ref\CDEO{B.A. Campbell, S. Davidson, J. Ellis and K.A. Olive,
 Phys. Lett. B256 (1991) 457.}\ since
they  are model dependent
\ref\CKO{J.M. Cline, K. Kainulainen and K.A. Olive,
 Phys. Rev. Lett. 71 (1993) 2372; \hfill\break
  Phys. Rev. D49 (1994)  6394.}
and may be evaded in some baryogenesis scenarios
\ref\ewbaryo{G. Farrar and M. Shaposhnikov,
 Phys. Rev. Lett. 70 (1993) 2833;\hfill\break
 Erratum {\it ibid.} 71 (1993) 210; CERN-TH-6732-93 (1993).}.)

These bounds pose a serious problem for generic SUSY models where
the natural expectation is that $\vev{\nu_i}\sim\vev{\phi_d}\sim m_Z$,
$\mu_{\nu_i\phi_u}\sim\mu_{\phi_d\phi_u}\sim m_Z$ and $\l_{ijk}\sim1$.
The standard solution to this problem is to impose
a discrete symmetry, $R$-parity ($R_p$), that forbids all three
types of terms. Alternatively, one could just impose lepton
number to forbid these terms.

In this work, we would like to suggest an alternative mechanism
to suppress SUSY contributions to neutrino masses: an approximate
alignment of the $\mu$ terms and the SUSY violating $B$ terms.
(Hall and Suzuki \HaSu\ noted this
case parenthetically in their study of models without R parity
but did not emphasize it because it did not fit into the
Grand Unified framework  which was one of their primary concerns.)
We will first present the mechanism and then show that it arises
naturally in the framework of abelian horizontal symmetries.
Furthermore, such symmetries automatically suppress the trilinear
lepton number violating couplings.

\newsec{Alignment}
In this section, we introduce our notations, clarify the
meaning of the bounds \boundVEV\ and \boundmu\ and present
a mechanism that may satisfy these bounds.

\subsec{Notations}
In Supersymmetric extensions of the Standard Model without
$R_p$ or lepton number, there is a-priori nothing to distinguish
the lepton-doublet supermultiplets $L_i$ from the down-Higgs
supermultiplet $\phi_d$, as both transform as $(2)_{-1/2}$ under
$SU(2)_L\times U(1)_Y$. We denote then the four $Y=-1/2$ doublets
as $L_\alpha$, $\alpha=0,1,2,3$. The single $\mu$-term of the
Minimal Supersymmetric Standard Model
(MSSM) is now extended to a four-vector,
\eqn\mudef{\mu_\alpha L_\alpha\phi_u,}
where $\phi_u(2)_{+1/2}$ is the up-Higgs supermultiplet.
The single SUSY breaking $B$ term of the MSSM
is also extended to a four-vector,
\eqn\Bdef{B_\alpha L_\alpha\phi_u,}
where here $L_\alpha$ and $\phi_u$ stand for the scalar components
in the supermultiplets. The trilinear terms in the superpotential
contain lepton number violating generalizations of the down quark
and charged lepton Yukawa matrices,
\eqn\lamdef{\l_{\alpha\beta k}L_\alpha L_\beta\bell_k+\lp_{\alpha jk}
L_\alpha Q_j\bar d_k,}
where $\bell_k(1)_{+1}$ are the three lepton singlets,
$Q_j$ are quark doublets and $\bar d_k$ are down quark singlets.
Finally, there are also SUSY breaking scalar masses,
\eqn\msqdef{m^2_{\alpha\beta}L_\alpha^\dagger L_\beta+{\rm h.c.}}
that are relevant to our study. (Here, again, $L_\alpha$ stand for the
scalar components.)

\subsec{Neutralinos}
The full neutralino mass matrix is $7\times7$, with rows and columns
corresponding to
$\{\tilde\gamma,\tilde z,\tilde\phi_u^0,\tilde L_\alpha^0\}$.
(Here, $\tilde L_\alpha$ corresponds to the fermionic components
in $L_\alpha$.) Neutrino masses arise from the $6\times6$
mass matrix $M^{\rm n}$ (the photino is irrelevant to neutrino masses)
\eqn\Mneu{M^{\rm n}=\pmatrix{
m_{\tilde z}&-{g\over2\cos\theta_W}v_u&{g\over2\cos\theta_W}v_\alpha\cr
-{g\over2\cos\theta_W}v_u&0&\mu_\alpha\cr
{g\over2\cos\theta_W}v_\alpha&\mu_\alpha&0_{4\times4}\cr},}
where $v_u=\vev{\phi_u^0}$, $v_\alpha=\vev{L_\alpha^0}$, and
$0_{4\times4}$ denotes a zero $4\times4$ block in $M^{\rm n}$.
(The zeros in this $4\times4$ block are lifted by non-renormalizable
terms in the superpotential, {\it i.e.} ${1\over M}\phi_u\phi_u LL$.
Taking $M\gg m_Z$, these terms have negligible effects on our
discussion. Here and in our analysis below
we neglect these terms as well as  additional small loop effects.)
In general, $M^{\rm n}$ gives 4 massive states and two massless ones.
Three of the four massive states should correspond to (combinations of)
the zino and the two Higgsinos with masses $\sim\O(m_Z)$.
The remaining massive state is then one of the neutrinos, and its mass is
constrained by \numass.

The product of the four masses is easily extracted from \Mneu. Define
\eqn\defmuvev{\eqalign{
\mu\equiv&\ (\sum_\alpha\mu_\alpha^2)^{1/2},\cr
v_d\equiv&\ (\sum_\alpha v_\alpha^2)^{1/2},\cr
\cos\xi\equiv&\ {\sum_\alpha v_\alpha\mu_\alpha\over v_d\mu}.\cr}}
Note that $\xi$ measures the alignment of $v_\alpha$ and $\mu_\alpha$.
We find
\eqn\detmassn{{\det}^\prime M^{\rm n}\sim\mu^2 v_d^2\sin^2\xi,}
where by $\det^\prime$ we mean the product of (in our case, the four)
eigenvalues different from zero.
Following the discussion above, we require
\eqn\numassdet{\mu^2 v_d^2\sin^2\xi\lsim m_Z^3 m_\nt,}
where $m_\nt$ stands for the heaviest among the neutrino
mass eigenstates: $m_\nt\lsim24\ MeV$ from direct experiments
or $m_\nt\lsim10\ eV$ from cosmology if its lifetime is longer
than the age of the Universe. Note, however, that $\mu\sim\O(m_Z)$
because it provides the charged Higgsino masses, and (for $\tan\beta\sim
1$) $v_d\sim\O(m_Z)$ because it contributes sizably to $m_Z$ and $m_W$
and it provides the down quark and charged lepton masses.
The requirement is then
\eqn\alignvdmu{\sin\xi\lsim\O\left(\sqrt{m_\nt\over m_Z}\right).}

The bound \numassdet\ or, equivalently, \alignvdmu\ is a severe
constraint on SUSY models because generically one expects
$\sin\xi\sim\O(1)$. It translates into \boundVEV\ and \boundmu\
in the following way: take $v_\alpha$ and $\mu_\alpha$ to be
approximately aligned. Then there are three mass eigenstates
of $M^{\rm n}$ with masses of $\O(m_Z)$. Eq. \numassdet\ implies
that the vev $\vev{L_\alpha}$ in the direction orthogonal
to these three massive states should be $\lsim\O(\sqrt{m_Z m_\nt})$
(eq. \boundVEV) and, similarly, $\mu_\alpha$ in this direction
should be $\lsim\O(\sqrt{m_Z m_\nt})$ (eq. \boundmu).

To summarize, in phenomenologically consistent models,
both $\mu$ and $v_d$ are of $\O(m_Z)$ and
approximately aligned; the misalignment should not exceed
$\O(10^{-2})$ or even $\O(10^{-5})$ if the cosmological bound holds.

\subsec{Charginos}
In the previous sub-section, we have shown that out of the seven
neutral fermions, two are (to the approximation in which we work)
massless but, in general,
five are massive. To guarantee a third very light (compared to the
electroweak breaking scale) neutral fermion, $v_\alpha$ and $\mu_\alpha$
have to be aligned. There is still a question, however, of whether
the three resulting light states correspond to neutrinos. To answer
this question, we have to study the charged fermion mass matrix.

The chargino mass matrix $M^{\rm c}$ is $5\times5$, with
rows corresponding to $\{\tilde w^-,\tilde L_\alpha^-\}$, and columns to
$\{\tilde w^+,\tilde\phi_u^+,\bell_k^+\}$:
\eqn\Mneu{M^{\rm c}=\pmatrix{
M_2&{g\over\sqrt2}v_u&0_{1\times3}\cr
{g\over\sqrt2}v_\alpha&\mu_\alpha&\l_{\alpha\beta k}v_\beta\cr}.}
Note that the $SU(2)_L$ gauge symmetry implies that
$\l_{\alpha\beta k}$ is antisymmetric in $(\alpha,\beta)$ and,
therefore, $(M^{\rm c})_{\alpha k} v_\alpha=0$.

Let us now assume that the phenomenological constraint \alignvdmu\
is fulfilled, namely $v_\alpha$ and $\mu_\alpha$ are approximately
aligned. Then, to a very good approximation,
$(M^{\rm c})_{\alpha k}\mu_\alpha=0$. To understand the consequences,
it is convenient to define
\eqn\redefL{\phi_d={1\over v_d}\sum_\alpha v_\alpha L_\alpha,}
and $L_i$ as the three fields orthogonal to $\phi_d$. The charged fermion
mass matrix with rows corresponding to
$\{\tilde w^-,\tilde\phi_d^-,\tilde L_i^-\}$
(and columns as above) is, to a very good approximation, block-diagonal:
\eqn\redefM{M^{{\rm c}\prime}=
\pmatrix{M_2&{g\over\sqrt2}v_u&0_{1\times3}\cr
{g\over\sqrt2}v_d&\mu&0_{1\times3}\cr 0_{3\times1}&0_{3\times1}&
\l_{i \phi_d k} v_d\cr}.}
(The zeros in the second column stand for highly suppressed entries,
of order $\mu\sin\xi$; the other zeros are exact for renormalizable
tree-level terms.) We learn the following:
\item{(i)} The three singlets $\bell_i$ do not mix, to
a good approximation, with the triplet $\tilde w$ and doublet
$\tilde\phi_u$. This implies that the mass eigenstates,
whose right handed components are $\bell_i$, are the `charged leptons'.
\item{(ii)} The left handed components in the charged leptons
come from the three $L_i$.
\item{(iii)} Neutrinos, which are defined as the $SU(2)_L$ partners
of the left handed charged leptons, correspond then to the
three neutral members in $L_i$.

However, our analysis of the neutralino mass matrix reveals that,
for $\mu_\alpha\propto v_\alpha$, the three neutral fermion components in
$L_i$ correspond to the three light mass eigenstates. We conclude
that aligning $\mu_\alpha$ with the VEV $v_\alpha$ guarantees not only
that there are three very light neutral fermions, but also that these
light states are the three neutrinos.

\subsec{Alignment}
The alignment of $\mu_\alpha$ with $v_\alpha$,
\eqn\vevalign{\mu_\alpha\propto v_\alpha,}
can be achieved by imposing two conditions on the SUSY parameters:
\item{(a)} The $B$-terms are proportional to the $\mu$-terms \HaSu:
\eqn\prop{B_\alpha\propto\mu_\alpha.}
\item{(b)} $\mu_\alpha$ is an eigenvector of $m^2_{\alpha\beta}$
(the SUSY-breaking scalar mass-squared matrix):
\eqn\univ{m^2_{\alpha\beta}\mu_\beta=\tilde m^2 \mu_\alpha.}

To prove this statement, note that the minimum
equations that determine $v_\alpha$ depend on $\mu_\alpha$,
$B_\alpha$, $m^2_{\alpha\beta}$ and gauge couplings.
In particular, the minimum equations do not depend on
the trilinear couplings $\l_{\alpha\beta k}$ and $\lp_{\alpha jk}$,
because these always involve a charged field. It is convenient to
rotate to a basis where $m^2_{\alpha\beta}$ is diagonal.
Condition (b) guarantees that, in this basis, $\mu_\alpha$ has only
a single component, say $\mu_0$, that is different from zero.
Condition (a) guarantees that also $B_\alpha$ has only $B_0\neq0$.
Then trivially (in similarity to the R parity case) $\vev{L_0}\neq0$,
$\vev{L_i}=0$, is a solution of the minimum equations,
namely \vevalign\ holds.

We conclude that when \prop\ and \univ\ hold, neutrinos do not mix
with gauginos and Higgsinos
and their masses are, therefore, highly suppressed.

One could think of various theoretical frameworks where
\prop\ and \univ\ hold. For example, if string theory
guarantees that the $\mu$ terms arise from the K\"ahler
potential only and if the quadratic terms in the K\"ahler
potential depend weakly on the moduli whose
$F$-terms break supersymmetry, then $B$ and $\mu$ would be
approximately aligned. However, in this work we would like
to show that the required alignment arises naturally
in the framework of horizontal symmetries.

\newsec{Horizontal Symmetries}
 The hierarchical pattern of fermion masses and mixing angles
could be the result of an abelian horizontal symmetry that is
explicitly broken by a small parameter.  With a single breaking
parameter $\l$, whose charge under the horizontal symmetry is
defined to be $H(\l)=-1$, the following selection rules apply:
\item{a.} Terms in the superpotential that carry charge $n\geq0$ under
$\H$ are suppressed by $\O(\l^n)$, while those with $n<0$ are forbidden
due to the holomorphy of the superpotential. (If $\H=Z_N$, the
suppression is by $\O(\l^{n(\mod N)})$.)
\item{b.} Terms in the K\"ahler potential that carry charge $n$ under
$\H$ are suppressed by $\O(\l^{|n|})$ (or $\O(\l^{{\rm min}[\pm n(\mod
N)]})$ for $\H=Z_N$).

The selection rules apply to all orders in perturbation
theory, so we can safely ignore loop effects.

Note that the $\mu$-terms in the effective low-energy superpotential
could originate from either or both of the
high energy superpotential and the high energy K\"ahler potential.
The superpotential contributions obey rule a, and their scale is
arbitrary. Those from the K\"ahler potential obey rule b and their
natural scale is the SUSY breaking scale $\tm$
\ref\GiMa{G.F. Giudice and A. Masiero, Phys. Lett. B206 (1988) 480.}
\ref\KaLo{V.S. Kaplunovsky and J. Louis, Phys. Lett. B306 (1993) 269.}.

For the various terms relevant to our study, the following
order of magnitude estimates hold (we use $U(1)_Y$ to set $H(\phi_u)=0$):
\eqn\oom{\eqalign{
\mu_\alpha\sim&\ \cases{\mu^0\l^{H(L_\alpha)}&$H(L_\alpha)\geq0$,\cr
\tm\l^{|H(L_\alpha)|}&$H(L_\alpha)<0$,\cr}\cr
B_\alpha\sim&\ \tm B^0\l^{|H(L_\alpha)|},\cr
m^2_{\alpha\beta}\sim&\ \tm^2\l^{|H(L_\beta)-H(L_\alpha)|}.\cr}}
Here, $\mu^0$ and $B^0$ are unknown `natural' scales for $\mu$
and $B/\tm$, respectively, and $\tm$ is the SUSY breaking scale.
Eqs. \oom\ lead to the following simple observations:
\item{a.} Assuming that all $H(L_\alpha)$ are of the same sign and
that one of the $L_\alpha$ fields (say, $L_0$) carries the smallest
horizontal charge, $|H(L_0)|\ll|H(L_i)|$ ($i=1,2,3$), then both the
$\mu_\alpha$ terms and the $B_\alpha$ terms will be dominantly in
the direction of this field:
\eqn\Halign{\mu_0\gg\mu_i,\ \ \ B_0\gg B_i.}
\item{b.} The diagonal terms in $m^2_{\alpha\beta}$ are not
suppressed by the selection rules, namely $m^2_{\alpha\alpha}\sim\tm^2$,
while the off-diagonal are suppressed
if the various fields have different charges,
\eqn\Huniv{m^2_{\alpha\beta}\ll m^2_{\alpha\alpha}\sim\tm^2
\ \ \ (\alpha\neq\beta).}
The important point here is that
${m^2_{0i}\over m^2_{00}}\sim{\mu_i\over\mu_0}$.

These two effects fulfill the two conditions described in the
previous section in an {\it approximate} way. Consequently, the mixings
of neutrinos with the zino and the Higgsino do not vanish but
are suppressed. It now becomes a {\it quantitative} question
of whether reasonable horizontal charge assignments lead to
satisfactory suppression of neutrino masses.

Note that, since the mixing between $L_0$ and the three $L_i$ is
very small, we can neglect the rotation \redefL\ from the
$\{L_\alpha\}$ basis to the $\{\phi_d,L_i\}$ basis in our
various order of magnitude estimates.

The quantitative answer is easy to find: as $\sin\xi\sim
\O({\mu_i\over\mu_0})\sim\O({B_i\over B_0})\sim\O({m^2_{0i}\over
m^2_{00}})$, eq. \alignvdmu\ (or, equivalently,
\boundVEV\ and \boundmu) is satisfied if
\eqn\boundcharge{
\l^{H(L_\tau)-H(\phi_d)}\lsim\sqrt{m_\nt\over m_Z}\lsim\cases{
10^{-2}&$m_\nt\leq24\ MeV$,\cr 10^{-5}&$m_\nt\lsim10\ eV$.\cr}}
If the small parameter $\l\sim0.2$, as suggested by the magnitude of
the Cabibbo angle, then
\eqn\chargedif{H(L_\tau)-H(\phi_d)\gsim\cases{
3&$m_\nt\leq24\ MeV$,\cr 7&$m_\nt\lsim10\ eV$.\cr}}
A charge difference of $\O(7)$ may be too large for reasonable models.
However, in some models of ref.
\ref\lnsa{M. Leurer, Y. Nir and N. Seiberg,
 Nucl. Phys. B398 (1993) 319.},
where the symmetry breaking
parameters are much smaller than 0.2, the required
approximate alignment can be achieved with charge differences $\leq2$.

Eq. \chargedif\ ensures that, at tree level, $m_\nt$ is safely
suppressed. One may still worry whether loop corrections can give
larger contributions to the neutrino masses. However, this is not
the case. The leading contributions come from loops generated by
the $\lp_{ijk}$ couplings \lamdef\ with $d$-type quarks--squarks
circulating in the loop. They are proportional
to the heaviest $d$-quark mass $m_b$ :
$(\delta m_\nu)_{ij}\sim{1\over16\pi^2}\lp_{i3k}\lp_{jk3} m_b$. Since
$\lp_{i3k}\lsim\l^{H(L_i)-H(\phi_d)}{m_b\over m_Z}$,
the radiative contributions to $m_{\nu_i}$ require
\eqn\rad{
\l^{H(L_i)-H(\phi_d)}\lsim\sqrt{m_{\nu_i}\over m_Z}\  4\pi\
\left({m_Z\over m_b}\right)^{3/2}.}
This is weaker than \boundcharge\ by a factor $\sim10^3$.
Eq. \rad\ shows explicitly how the suppression from horizontal
symmetries is effective at any order in perturbation theory,
and indeed justifies neglecting loop effects.

We conclude that in models of abelian horizontal symmetries,
the $\mu$ and $B$ terms are dominantly in the direction of one of
the four $L_\alpha$ fields, and the scalar mass--squared matrix does
not significantly mix this field with the other three.
This leads to an approximate alignment of $\mu_\alpha$ and $v_\alpha$.
Consequently, neutrino masses from mixing
with the zino or Higgsino can be suppressed well below the electroweak
scale, while radiative contributions can be kept negligibly small.
Whether this suppression is strong enough is a model dependent
question. We present a class of models with satisfactory
suppression in the next section.

\newsec{An Explicit Example}

Take a model with an exact discrete horizontal symmetry
\eqn\Hsym{\H=Z_{n_1}\times Z_{n_2}.}
The symmetry is spontaneously broken -- as we show below -- by
two scalars in singlet supermultiplets:
\eqn\Srep{S_1(-1,0),\ \ \ S_2(0,-1),}
In addition, we have the doublet supermultiplets:
\eqn\prep{\phi_u(0,0),\ \ \ L_\alpha(H_{1\alpha},H_{2\alpha}).}
We use horizontal charges $0\leq H_{i\alpha}\leq n_i-1$.

In order to estimate the VEVs of the various fields, we investigate
the Higgs potential and the minimum equations. We assume that there
are only two scales in the model: $\tm$ is the SUSY breaking scale
which characterizes all SUSY breaking terms, and $M_p$, the Planck scale
which suppresses all non-renormalizable terms. We consider all the
terms that are consistent with $SU(2)_L\times U(1)_Y\times \H$.
We omit dimensionless coefficients of $\O(1)$ in all formulae.

The leading terms in the superpotential are
\eqn\Wtwo{W\sim{S_1^{n_1}\over M_p^{n_1-3}}+{S_2^{n_2}\over M_p^{n_2-3}}+
\sum_\alpha{S_1^\Ho S_2^\Ht(\phi_u L_\alpha)\over M_p^{\Ho+\Ht-1}}.}
They lead to the following (leading) terms in the Higgs potential:
\eqn\VW{\eqalign{V^W\sim&\ {|S_1|^{2n_1-2}\over M_p^{2n_1-6}}
+{|S_2|^{2n_2-2}\over M_p^{2n_2-6}}\cr
+&\ \sum_\alpha\left[\left({S_1^*\over M_p}\right)^{n_1-1}
{S_1^{\Ho-1}S_2^{\Ht}\over M_p^{\Ho+\Ht-3}}
+\left({S_2^*\over M_p}\right)^{n_2-1}
{S_1^{\Ho}S_2^{\Ht-1}\over M_p^{\Ho+\Ht-3}}\right](\phi_u L_\alpha)
+{\rm h.c.}.\cr}}
In addition, there are D terms in the Higgs potential,
\eqn\VD{V^D\sim(|\phi_u|^2-\sum_\alpha |L_\alpha|^2)^2,}
soft scalar masses (off-diagonal terms are highly suppressed),
\eqn\VA{V^A\sim
\tm^2(|S_1|^2+|S_2|^2+|\phi_u|^2+\sum_\alpha|L_\alpha|^2),}
and (the leading) soft SUSY breaking terms analytic in the fields,
\eqn\VB{V^B\sim\tm\left[
{S_1^{n_1}\over M_p^{n_1-3}}+{S_2^{n_2}\over M_p^{n_2-3}}+\sum_\alpha
{S_1^{\Ho}S_2^{\Ht}(\phi_u L_\alpha)\over M_p^{\Ho+\Ht-1}}
\right].}
In \VW, \VD, \VA\ and \VB, the various fields stand for the
neutral scalar components.

Solving the minimum equations for $\vev{S_i}$, we get the two small
breaking parameters for $H$:
\eqn\solveS{\l_1\equiv
{\vev{S_1}\over M_p}\sim\left({\tm\over M_p}\right)^{1\over n_1-2},
\ \ \ \l_2\equiv{\vev{S_2}\over M_p}\sim
\left({\tm\over M_p}\right)^{1\over n_2-2}.}
This is a generalization of the $\H=Z_n$ case studied in ref.
\ref\lnsb{M. Leurer, Y. Nir and N. Seiberg,
 Nucl. Phys. B420 (1994) 468.}.
For the scalar doublet VEVs, we get
\eqn\solvephi{{\vev{\phi_u}\over\tm}\sim1,}
\eqn\solveL{{\vev{L_\alpha}\over\tm}\sim\left({\tm\over M_p}\right)^
{{\Ho\over n_1-2}+{\Ht\over n_2-2}-1}.}

Equation \solveL\ shows that, indeed, the VEVs of the $L_\alpha$
doublets are hierarchical and depend on the horizontal charges
in the way described in the previous section. The effective
$\mu_\alpha$ can be extracted from eq. \Wtwo\ by putting in the
VEVs $\vev{S_i}$, with the result
$\mu_\alpha^{{\rm eff}}\sim\vev{L_\alpha}$. That is,
$\mu_\alpha$ and $\vev{L_\alpha}$ are approximately aligned.
Taking $\vev{L_0}>\vev{L_3}$ to be the two largest of the four
$\vev{L_\alpha}$, the alignment is accurate to order
$\l_1^{H_1(L_3)-H_1(L_0)}\l_2^{H_2(L_3)-H_2(L_0)}$.

The VEV of the down Higgs (for $\tan\beta\sim1$) should be of $\O(\tm)$.
This is achieved if one of the $L_\alpha$ fields, say $L_0$,
has one of its horizontal charges
$H_i(L_0)=n_i-2$ and the other $H_j(L_0)=0$:
\eqn\HHphid{L_0(n_1-2,0)\ \Longrightarrow\
\vev{L_0},\mu_0^{\rm eff}\sim\tm.}
If we then take, for example,
\eqn\HHLi{L_i(H_{1i},n_2-2)\ \Longrightarrow\ \vev{L_i},\mu_i
\sim\tm\left({\tm\over M_p}\right)^{{H_{1i}\over n_1-2}},}
so that
\eqn\chooseHo{{H_{1i}\over n_1-2}\gsim{1\over3}\ \Longrightarrow\
\vev{L_i},\mu_i\lsim10^{-5}\tm,}
the alignment is precise to $\O(10^{-5})$,
and neutrino masses are safely below the cosmological bound.

The model presented here, in addition to naturally suppressing
neutrino masses, has two more attractive features \lnsb:
\item{1.} Eq. \solveS\ shows that a hierarchy of VEVs that could be
relevant to fermion parameters
can arise naturally out of the initial two-scale model.
\item{2.} Eq. \HHphid\ shows that the horizontal symmetry can naturally
solve the $\mu$-problem.

The model may seem complicated, but the reason is that we want to
demonstrate the power of horizontal symmetries in naturally
achieving these extra advantages. A model with a gauged horizontal
$U(1)$ symmetry, with given small breaking parameters and
a given scale $\mu^0$, would achieve the required alignment
in lepton parameters with much simpler charge assignments.
(See, for example, the models of ref.
\ref\lmmm{Y. Grossman and Y. Nir, WIS-95/7/Mar-PH, hep-ph/9502418.}.)

We also note that a model without $R_p$ and with the $L_\alpha$
transforming non-trivially under a single horizontal $Z_n$
does not work. The VEVs of the doublet fields are
$\vev{L_\alpha}\sim\tm\left({\tm\over M_p}\right)^{{H_\alpha\over
n-2}-1}$. Consequently, for $H_\alpha<n-2$ (which is unavoidable for
some of the horizontal charges) $\vev{L_\alpha}>\tm$, and the
electroweak symmetry is broken at a scale higher than the
SUSY breaking scale. Of course, with $R_p$, models
with a single $Z_n$ and $H(\phi_d)=n-2$ do solve the $\mu$ problem \lnsb.

To demonstrate the full power of the discrete horizontal symmetry,
we suggest the following explicit example. Take $\H=Z_{14}\times Z_{10}$,
with $S_i$ of eq. \Srep, $\phi_u$ of eq. \prep, and
\eqn\explicitL{L_0(12,0),\ L_3(4,8),}
(and higher charges for $L_1$, $L_2$).
Solving the minimum equations and studying the neutrino spectrum,
we find:
\par (i) The two small breaking parameters are
\eqn\goodS{\l_1\equiv{\vev{S_1}\over M_p}\sim\l^2,\ \
\l_2\equiv{\vev{S_2}\over M_p}\sim\l^3,}
which could explain all quark and lepton parameters, as shown in refs.
\lnsa \lnsb \lmmm.
\par (ii) The $\mu$-problem is solved namely, identifying $\phi_d\sim
L_0$,
\eqn\goodphi{\vev{\phi_u}\sim\vev{\phi_d}\sim\mu_0\sim\tm.}
\par (iii) Neutrino masses are highly suppressed. In particular,
$m_\nt\sim10\ eV$ which is the relevant range for being hot dark matter.

\newsec{Trilinear Lepton Number Violating Terms}
We investigate the dimension-4 terms in the superpotential:
\eqn\dimfour{
\l_{ijk}L_iL_j \bell_k + \lp_{ijk}L_iQ_j \bar d_k +
\ld_{ijk} \bar u_i \bar d_j \bar d_k.}
(Non-renormalizable lepton number violating terms pose no problem.)
In our presentation below we neglect the rotation
{}from the interaction basis (where the horizontal charges are
well defined) to the mass basis: a full analysis, involving an estimate
of the mixing angles (which are also determined by the horizontal
charges), would give just the same order of magnitude estimates.

The selection rules, when applied to these couplings, imply
\eqn\selctl{\eqalign{
\l_{ijk}\sim&\ \cases{\l^{H(L_i)+H(L_j)+H(\bell_k)}&
$H(L_i)+H(L_j)+H(\bell_k)\geq0$\cr
0&$H(L_i)+H(L_j)+H(\bell_k)<0$\cr}\cr
\lp_{ijk}\sim&\ \cases{\l^{H(L_i)+H(Q_j)+H(\bar d_k)}&
$H(L_i)+H(Q_j)+H(\bar d_k)\geq0$\cr
0&$H(L_i)+H(Q_j)+H(\bar d_k)<0$\cr}.\cr}}
We first assume baryon number conservation (which requires
$\ld_{ijk}=0$). Stringent bounds apply to products of two $\l$'s
\HiKa
\ref\DH{S. Dimopoulos and L.J. Hall, Phys. Lett. B207 (1987) 210.}
\ref\DrRo{H. Dreiner and G.G. Ross, Nucl. Phys. B365 (1991) 597.}
\ref\BGMT{R. Barbieri {\it et. al.}, Phys. Lett B252 (1990) 251.}
\ref\Moh{R.N. Mohapatra, Prog. Part. Nucl. Phys. 31 (1993) 39.}.
These are given in Table I.

\vskip 1.0cm
\centerline{Table I: Constraints on Lepton Number Violating Couplings}
\vskip 0.5cm
\begintable
Couplings      | Limit      | Process      | Master model \cr
$\lp_{k12}\lp_{k21}$|$9\times10^{-8}\sim\l^{10}$|
$\Delta m_K$ | $\l^{10}$ \cr
$\lp_{k12}\lp_{k21}$ | $8 \times 10^{-10} \sim \l^{13}$ |
$\epsilon$  |  $\l^{10}$ \cr
$\lp_{k13}\lp_{k31}$ | $4 \times10^{-6} \sim \l^8$ |
$\Delta m_B$ | $\l^8$ \cr
$\l_{k12}\l_{k11}$   | $10^{-4} \sim \l^6$   |
$\mu \rightarrow e e e$ | $\l^{15}$ \cr
$\l_{2jk}\l_{1jk}$   | $10^{-3} \sim \l^5$|
$\mu \rightarrow e \gamma$ | $\l^{12}$ \cr
$\lp_{k21}\l_{k21}$ | $2 \times10^{-5} \sim \l^7$|
$K_L \rightarrow \mu e$ | $\l^{11}$ \cr
$\lp_{k21}\l_{k22}$ | $10^{-4}\sim \l^6$  |
$K_L \rightarrow \mu \mu$ | $\l^{10}$ \cr
$\lp_{k21}\l_{k11}$  | $2 \times10^{-5} \sim \l^7$|
$K_L \rightarrow e e$ | $\l^{13}$ \endtable

Note the following:

(a) All bounds correspond to $\tm=1\,TeV$ and scale like $1/\tm^2$.

(b) The bounds from $K_L$ decays also apply to
$\l_{k21}\ra\l_{k12}$, $\lp_{k21}\ra\lp_{k12}$. In all these
other cases the horizontal symmetry
gives similar or even stronger suppression.

(c) We do not present various additional bounds that require
$\l_{ijk},\ \lp_{ijk}\lsim\l^2$ and are easily satisfied
in any of our models.

(d) The master model for quarks was presented in ref. \lnsb.
For the lepton sector, we assume \lmmm\
\eqn\leptonmaster{
V_{e\mu}\sim\l^2,\ \ {m_\nm\over m_\nt}\sim\lambda^4,\ \
{m_e\over m_\mu}\sim\lambda^3,\ \  {m_\mu\over m_\tau}\sim\lambda^2.}

The conclusion is that all the bounds are satisfied within our
horizontal symmetry models. The only potential problem is in
$\epsilon$ if we assume phases of $\O(1)$. This can be solved by a slight
modification of the `master' model: Choosing horizontal charges
$H^\prime=H+\alpha L$ (where $L$ is lepton number and $\alpha$ is
a real coefficient), one can achieve an arbitrary suppression of the
lepton number violating terms in \dimfour, while the only effect on the
fermion mass matrices is an overall suppression of all neutrino masses.

Baryon number violation was investigated in ref.
\ref\BHNi{V. Ben-Hamo and Y. Nir, Phys. Lett. B339 (1994) 77.}\
assuming massless neutrinos. With slight modifications of their
models, a satisfactory suppression of proton decay can be achieved
for the massive neutrino case as well.

To summarize:
Assuming baryon number conservation, dimension 4 lepton number violating
terms are suppressed to a phenomenologically acceptable level by a
horizontal  symmetry.  In ref. \BHNi\ models were constructed
in which horizontal
symmetry rather than baryon number suppresses proton decay and
other B violating processes.  Simple modifications of
those models lead to a horizontal symmetry framework in which all of
the usual phenomenological consequences
of baryon number, lepton number and R parity follow.

\newsec{Conclusions}
Supersymmetric models without R parity and without lepton number
symmetry lead, in general, to an unacceptably large neutrino mass.
This problem is solved, however, in any model where (similarly
to models with R parity), the vacuum expectation value of the four
$Y=-1/2$ doublet scalars is aligned with the $\mu$ term which couples
these fields to the $Y=+1/2$ doublet scalar. For this alignment
to arise, two conditions have to hold: the soft SUSY breaking
$B$ term is proportional to the $\mu$ term, and the $\mu$ term is
an eigenvector of the SUSY breaking
scalar masses of the $Y=-1/2$ doublet scalars.

Models of abelian horizontal symmetries, with charges dictated
by fermion masses and mixing, automatically fulfill these
conditions but in an approximate way. The resulting approximate
alignment could lead to satisfactorily small neutrino masses.
In addition, trilinear lepton number violating terms in the
superpotential are allowed but suppressed below experimental
constraints. The resulting phenomenology could differ
significantly from models with exactly conserved  R parity in low energy
processes \BGH\Moh\BGMT
\ref\Moha{R. Mohapatra, Phys. Rev. D34 (1986) 3457.}
\ref\BaMa{R. Barbieri and A. Masiero, Nucl. Phys. B267 (1986) 679.},
in collider experiments
\DH \DrRo
\ref\BBHH{R. Barbieri, D.E. Brahm, L.J. Hall and S.D.H. Hsu,
 Phys. Lett. B238 (1990) 86.}
\ref\LoMc{S. Lola and J. McCurry, Nucl. Phys. B381 (1992) 559.}
\ref\GRV{M.C. Gonzales-Garcia, J.C. Romao and J.W.F. Valle,
 Nucl. Phys. B391 (1992) 100.}
\ref\Roy{ D.P. Roy, TIFR/TH/93-14, hep-ph/9303324.}
\ref\RRV{J.C. Romao, J. Rosiek and J.W.F. Valle, FTUV-95-02,
hep-ph/9502211.},
in the cosmological consequences
\ref\BS{A. Bouquet and P. Salati, Nucl. Phys. B284 (1987) 577.}
\ref\DiHa{S. Dimopoulos and L. Hall, Phys. Lett. B196 (1987) 135.}
\ref\BMV{V. Berezinskii, A. Masiero and J.W.F. Valle,
 Phys. Lett. B266 (1991) 382.}
and in some more peculiar effects, {\it e.g.}
matter enhanced neutrino oscillations
\ref\Roulet{E. Roulet, Phys. Rev. D44 (1991) 935.}
\ref\GMP{M.M. Guzzo {\it et. al.}, Phys. Lett B260 (1991) 154.}
\ref\BPW{V. Barger, R.J.N. Phillips and K. Whisnant,
 Phys. Rev. D44 (1991) 1629.}.
Most prominently, in the present framework, there is no reason for
the existence of a stable LSP.  This will change both the cosmological
and laboratory signals for supersymmetry.

\vskip 2cm
{\bf Acknowledgments}:
TB is a J.S. Guggenheim Fellow 1994-95 and
Varon Visiting Professor at the Weizmann Institute, and is
supported in part by DOE under grant DE-FG05-90ER40559.
YN is supported in part by the United States -- Israel Binational
Science Foundation (BSF), by the Israel Commission for Basic Research
and by the Minerva Foundation.

\listrefs
\end